\documentclass[12pt,a4paper]{article}
\usepackage{hyperref}
\usepackage[warn]{mathtext}
\usepackage[utf8]{inputenc}
\usepackage[T2A]{fontenc}
\usepackage[english]{babel}
\usepackage[titles]{tocloft}
\usepackage[a4paper]{geometry}

\usepackage{amsmath,amssymb,amsthm}
\usepackage[shortlabels]{enumitem}
\usepackage{braket}
\usepackage{graphicx}
\usepackage{float}
\usepackage{tikz}
\usepackage{standalone}
\usepackage{subcaption}

\usepackage{geometry}
\usepackage{authblk}

\providecommand{\Name}[1]{#1}
\providecommand{\Review}[1]{#1}
\providecommand{\Book}[1]{#1}
\providecommand{\Editor}[1]{#1}
\providecommand{\Publ}[1]{#1}
\providecommand{\Vol}[1]{\textbf{#1}}
\providecommand{\Year}[1]{(#1)}
\providecommand{\Page}[1]{#1}
\providecommand{\Pages}[2]{#1--#2}

\newcommand{\sx}{\sigma^x}
\newcommand{\sy}{\sigma^y}
\newcommand{\sz}{\sigma^z}
\newcommand{\ax}{\alpha_x}
\newcommand{\ay}{\alpha_y}
\newcommand{\az}{\alpha_z}
\newcommand{\msxn}{\langle \sx_0\rangle}
\newcommand{\msyn}{\langle \sy_0\rangle}
\newcommand{\mszn}{\langle \sz_0\rangle}
\renewcommand{\DH}{\Delta H}

\newcommand{\mean}[1]{\langle #1 \rangle}

\title{Quantifying the properties of evolutionary quantum states of the $XXZ$ spin model using quantum computing}

\author[1]{M. P. Tonne}
\author[1]{Kh. P. Gnatenko}

\affil[1]{Ivan Franko National University of Lviv, Professor Ivan Vakarchuk Department for Theoretical Physics - 12 Drahomanov St., Lviv, 79005, Ukraine}

\begin{document}

\maketitle
\abstract{
The entanglement distance of evolutionary quantum states of a two-spin system with the $XXZ$ model has been studied. The analysis has been conducted both analytically and using quantum computing. An analytical dependence of the entanglement distance on the values of the model coupling constants and the parameters of the initial states has been obtained. The speed of evolution of a two-spin system has been investigated. The analysis has been performed analytically and using quantum computing. An explicit dependence of the speed of evolution  on the coupling constants and on the parameters of the initial state has been obtained. The results of quantum computations are in good agreement with the theoretical predictions.}

\section{Introduction}

Compound quantum systems exhibit a unique property known as entanglement, which corresponds to non-classical correlations between the subsystems, where the state of a system cannot be factorized \cite{Einstein1935, Feynman1982, Horodecki2009}. Recent research has increasingly focused on entanglement and its applications \cite{Shimoni1995, Scott2004, Horodecki2002, Sheng2015,  Lloyd1996, Buluta2009, Huang2020, Briegel2001, Behera2019, GnatenkoSusulovska2022, Mooney2019}. It allows us to implement quantum algorithms that have no classical counterpart and to reach quantum supremacy \cite{Shor1997, Montanaro2016, Cerezo2021}.
Various entanglement measures have been studied. For example, \cite{Wang2018} employed negativity and computationally intensive state tomography to detect full entanglement in ring graph states. In other studies \cite{Kuzmak2020PL, KuzmakTkachuk2020, Susulovska2025}, entanglement was quantified as the distance between a target entangled state and the closest separable state.  In \cite{Cochiarella2020, Vesperini2023}, analogous geometric methods were applied to develop a distance-based entanglement measure for hybrid qudit systems. 

When estimating entanglement on a quantum computer, it is helpful to use a measure that relates to an observable that is easy to access. Authors in \cite{DefEnt} showed that the geometric measure of entanglement of a spin $1/2$, which represents a qubit, can be obtained from the mean value of this spin when it is coupled to any pure quantum state $\ket{\psi}$. 

The measure of entanglement known as entanglement distance was first defined in \cite{Cochiarella2020}.
It reads
\begin{equation}
    E_l^{ED}(\ket{\psi}) = 1 - \sum_{k=x,y,z} \langle\psi|\sigma_l^k|\psi\rangle^2,
\end{equation}
where $E^{ED}_l(\ket{\psi})$ is the entanglement of qubit $q[l]$ with other qubits in quantum state $\ket{\psi}$, $\sigma^k_l$,
$k = x, y, z$ are Pauli matrices corresponding to qubit (spin) $q[l]$ (see, for example, \cite{Vesperinni2024} and references therein).

In this paper, we use the entanglement distance as a measure of entanglement in the analysis of the system under consideration. The analysis is conducted both analytically and using quantum
computing. Quantum protocols for quantifying the mean values of spins are constructed. Protocols were run on the AerSimulator \cite{Aer}.

The geometric concept also makes it possible to identify another quantity of interest, namely, the speed of evolution of the system.
The speed of quantum evolution is defined as
\begin{equation}
    v = \lim_{\Delta \to 0} \frac{\Delta s}{\Delta t} = \frac{ds}{dt},
\end{equation}
where $\Delta s$ is the distance between two quantum states $\ket{\psi(t)}$, $\ket{\psi(t + \Delta t)}$. It is related to the quadratic fluctuations of energy. In addition, it is important to note that there exist several definitions of distance measures between two quantum states \cite{FPQM}, but for the quantum states that are close enough to each other (since $\Delta t \to 0$), all of them lead to the same expression for the speed of evolution. Namely, the speed of evolution is given by
\begin{equation}
    v = \frac{\gamma\sqrt{\langle \Delta H ^2 \rangle}}{\hbar},
    \label{eq:speed}
\end{equation}
where $\gamma$ is a constant (see \cite{Anandan}).
In papers \cite{GnatenkoSpeed2022}, \cite{GnatenkoSpeed2024}, the speed of evolution of quantum states of spin systems with the Ising model was quantified using quantum programming.
The evolution of qubits in time-dependent magnetic fields was analyzed in \cite{Cafaro2025}. Geometric quantum speed limits were investigated in open quantum systems in \cite{Pires2024}.
In this paper, we examine the speed of evolution of quantum states in a spin system with the XXZ model. The analysis is conducted both analytically and using quantum computing. Quantum protocols for quantifying the speed of evolution are constructed.
We run these protocols on the AerSimulator \cite{Aer}.

The paper is organized as follows. In Section 2, we analytically derive an expression that describes the entanglement distance in states generated by the operator of evolution with an anisotropic Heisenberg Hamiltonian. In Section 3, we have quantified the entanglement distance through quantum simulations on the AerSimulator.
In Section 4, we examine the speed of evolution in a two-spin system with the XXZ model analytically as well with quantum calculations.
The conclusions are highlighted in Section 5.

\section{Analytical consideration of entanglement distance of quantum states of a spin system with XXZ model}

Let us study the states generated by the evolution operator with an anisotropic Heisenberg Hamiltonian. We consider a system of two spins described by the Hamiltonian
\begin{equation}
    H = J \left(\sx_0\sx_1 + \sy_0\sy_1 + d\sz_0\sz_1 \right).
    \label{eq:Hamiltonian}
\end{equation}

The evolution operator corresponding to the Hamiltonian of the two-spin system takes the following form:
\begin{equation}   
    \hat{S} = exp \left(\frac{-i J \left(\sx_0\sx_1 + \sy_0\sy_1 + d\sz_0\sz_1 \right) t}{\hbar} \right).
\end{equation}

Taking into account that such operator pairs commute with each other $ \left[\sigma^k_0\sigma^k_1, \sigma^l_0\sigma^l_1\right] = 0$, where $k, l \in \{x, y, z\}$,  in the case of a two-qubit system, we can write
\begin{equation}
    \ket{\psi(t)} = RXX(\ax)RYY(\ay)RZZ(\az)\ket{\psi_0}
    \label{eq:evolution_state}
\end{equation}
The parameters $\ax = {2 J t}/{\hbar}$, $\ay = {2 J t}/{\hbar}$, $\az = {2 J d t}/{\hbar}$ are defined here for notational convenience and to reduce the complexity of expressions.

Let us consider the initial state to be  as follows
\begin{multline}
    \ket{\psi_0} =\left(\cos\frac{\theta_0}{2}\ket{0}_0 + e^{i\phi_0} \sin\frac{\theta_0}{2} \ket{1}_0 \right) \left(\cos\frac{\theta_1}{2}\ket{0}_1 + e^{i\phi_1} \sin\frac{\theta_1}{2} \ket{1}_1 \right)\\
    =e^{i\frac{\phi_0 + \phi_1}{2}}RZ_0(\phi_0)RZ_1(\phi_1)RY_0(\theta_0)RY_1(\theta_1) \ket{00} 
    \label{eq:init_state}
\end{multline}

Where $RZ_k(\phi_k) = exp(-i\phi_k\sz_k /2)$,
$ RY_k(\theta_k) = exp(-i\theta_k\sy_k / 2)$ are rotational gates acting on the states of the qubit (spin) $q[k]$. The expression above represents the most general form of a two-qubit non-entangled quantum state. The global phase does not affect any physical predictions and can be ignored.

Let us find the entanglement of the state $\ket{\psi(t)}$. For this purpose, we calculate $\msxn$, $\msyn$, $\mszn$ in the state $\ket{\psi(t)}$.
\begin{equation}
    \langle \sigma^k_0 \rangle = \bra{\psi_0} \hat{S}^+ \sigma^k_0 \hat{S} \ket{\psi_0}, 
\end{equation}
where $k \in \{x, y, z\}$.

For $\mszn$ we have
\begin{multline}
    \mszn = \bra{\psi_0} exp(i\ax\sx_0 \sx_1) exp(i\ay\sy_0 \sy_1)\sz\ket{\psi_0} \\
    = \cos(2\alpha_x) \cos(2\alpha_y) \cos(\theta_0)
    + \sin(2\alpha_x) \sin(2\alpha_y) \cos(\theta_1) \\
    + \frac{1}{4}\sin(2\alpha_x + 2\alpha_y) \sin\theta_0 \sin\theta_1 \sin\phi_0 \cos\phi_1,
    \label{eq:mean_z}
\end{multline}
here we take into account that $\sz_0$ anticommutes with $\sy_0$ and $\sx_0$, and commutes with $\sz_0$ and $\sigma^k_1$, therefore
\begin{equation}
     \sz_0 e^{\frac{-i J t \sx_0\sx_1 }{\hbar}} e^{\frac{-i J t \sy_0\sy_1 }{\hbar}} e^{\frac{-i J d t \sz_0\sz_1 }{\hbar}} 
     = e^{\frac{i J d t \sz_0\sz_1 }{\hbar}}e^{\frac{i J t \sy_0\sy_1 }{\hbar}} e^{\frac{-i J t \sx_0\sx_1 } {\hbar}} \sz_0.
\end{equation}

Analogously, for $\msxn$ and $\msyn$

\begin{multline}
    \msxn = \bra{\psi_0} e^{i\ax\sx_0 \sx_1} e^{i\ay\sy_0 \sy_1}\sz\ket{\psi_0} \\
    = 2 \sin\theta_0 \cos(2\alpha_y) \Bigl( \cos(\phi_0 + 2\alpha_z) \cos^2\frac{\theta_1}{2}
    + \cos(\phi_0 - 2\alpha_z) \sin^2\frac{\theta_1}{2} \Bigl) \\
    + 2 \sin\theta_1\sin(2\alpha_y) \Bigl(\sin(\phi_0 + 2\alpha_z) \cos^2\frac{\theta_0}{2} 
    - \sin(\phi_1 - 2\alpha_z) \sin^2\frac{\theta_0}{2}\Bigl)
    \label{eq:mean_x}
\end{multline}

\begin{multline}
    \msyn = \bra{\psi_0} e^{i\ax\sx_0 \sx_1} e^{i\ay\sy_0 \sy_1}\sz\ket{\psi_0} \\ 
    = 2 \sin\theta_0 \cos(2\alpha_x) \Bigl( \sin(\phi_0 + 2\alpha_z) \cos^2\frac{\theta_1}{2} 
    + \sin(\phi_0 - 2\alpha_z) \sin^2\frac{\theta_1}{2} \Bigl) \\
    + 2\sin\theta_1 \sin(2\alpha_x) \Bigl( -\cos(\phi_1 + 2\alpha_z) \cos^2\frac{\theta_0}{2} 
    + \cos(\phi_1 - 2\alpha_z) \sin^2\frac{\theta_0}{2} \Bigl).
    \label{eq:mean_y}
\end{multline}

So, the analytical expression for entanglement distance can be written as

\begin{multline}
    E = 1 - \Biggl(
    \biggl[
        2 \sin\theta_0 \cos(2\alpha_y) \Bigl(\cos(\phi_0 + 2\alpha_z) \cos^2\frac{\theta_1}{2} 
        + \cos(\phi_0 - 2\alpha_z) \sin^2\frac{\theta_1}{2} \Bigr) \\
        + 2 \sin\theta_1\sin(2\alpha_y) \Bigl(\sin(\phi_0 + 2\alpha_z) \cos^2\frac{\theta_0}{2} 
        - \sin(\phi_1 - 2\alpha_z) \sin^2\frac{\theta_0}{2} \Bigr)
    \biggr]^2  \\
    +\biggl[
        2 \sin\theta_0 \cos(2\alpha_x) \Bigl( \sin(\phi_0 + 2\alpha_z) \cos^2\frac{\theta_1}{2} 
        + \sin(\phi_0 - 2\alpha_z) \sin^2\frac{\theta_1}{2} \Bigr)\\
        + 2\sin\theta_1 \sin(2\alpha_x) \Bigl( -\cos(\phi_1 + 2\alpha_z) \cos^2\frac{\theta_0}{2} 
        + \cos(\phi_1 - 2\alpha_z) \sin^2\frac{\theta_0}{2} \Bigr)
    \biggr]^2 \\
    + \biggl[
        \cos(2\alpha_x) \cos(2\alpha_y) \cos(\theta_0) + \sin(2\alpha_x) \sin(2\alpha_y) \cos(\theta_1)\\
        + \frac{1}{4}\sin(2\alpha_x + 2\alpha_y) \sin\theta_0 \sin\theta_1 \sin\phi_0 \cos\phi_1
    \biggr]^2
    \Biggr)
    \label{eq:ent_dist}
\end{multline}

In the next section, we construct quantum protocols for quantifying entanglement and present the results of quantum computations.
% ===============================================================
\section{Quantifying the entanglement distance of quantum state with quantum calculations}

According to \cite{DefEnt}, to detect the entanglement of a quantum state, the observable value, namely the mean value of spin, has to be quantified. The mean value of the $\sz$ operator can be found directly based on the results of the measurements on the standard basis. That is, we have $\bra{\psi} \sz \ket{\psi}$
\begin{equation}
    \bra{\psi} \sz \ket{\psi} = |\langle \psi| 0 \rangle|^2 - |\langle \psi| 1 \rangle|^2
\end{equation}
To find the mean values of  $\sx$, $\sy$, and $\sz$ in a state $\ket{\psi}$ the following identities can be used
\begin{equation}
    \sx = exp(i\pi\sy /4)\sz exp(-i\pi\sy /4)
    \label{eq:sx_identity}
\end{equation}
\begin{equation}
    \sy = exp(i\pi\sx /4)\sz exp(-i\pi\sx /4)
    \label{eq:sy_identity}
\end{equation}
Thus, for the mean value of the operator $\sx$, we can write
\begin{equation}
    \bra{\psi}\sx\ket{\psi} = \bra{\psi^y} \sz \ket{\psi^y} = |\bra{\psi^y}0\rangle|^2 - |\bra{\psi^y}1\rangle|^2
\end{equation}
where we employ this notation $\ket{\psi^y} = exp(-i\pi\sy/4)\ket{\psi} = RY (-\pi/2) \ket{\psi}$.

Analogously, for $\sy$ we can write 
\begin{equation}
    \bra{\psi}\sy\ket{\psi} = \bra{\psi^yx} \sz \ket{\psi^x} = |\bra{\psi^x}0\rangle|^2 - |\bra{\psi^x}1\rangle|^2
\end{equation}
where $\ket{\psi^x} = exp(i\pi\sx /4)\ket{\psi} = RX (\pi/2) \ket{\psi}$.

Thus, summarizing, to obtain the expectation values of the Pauli operators, we measure in the computational basis after applying appropriate basis rotations. Since $\sz$ has eigenstates $\ket{0}$ and $\ket{1}$ with eigenvalues $\pm 1$, the mean value $\bra{\psi} \sz \ket{\psi}$ is simply the difference between the probabilities of obtaining these outcomes. The x- and y-components can be mapped to the z-basis through suitable rotations. Using the identities \eqref{eq:sx_identity} and \eqref{eq:sy_identity}, we see that measuring $\sz$ corresponds to rotating the qubit about the Y axis by $-\pi/2$, while measuring $\sy$ corresponds to rotating it about the X axis by $\pi/2$. Thus, to determine $\sx$ or $\sy$, one prepares the state, applies the corresponding rotation, and then performs a standard measurement in the computational basis. According to these transformations, the expectation values follow directly from the measurement statistics.

\begin{figure}
\centering
    \includegraphics[width=0.7\linewidth]{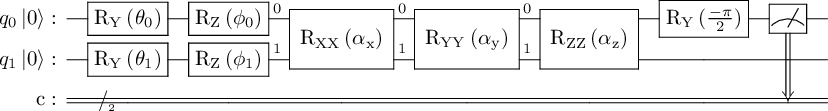}
    \caption{Quantum protocol for calculating the mean value of operator $\sx$}
    \label{fig:protocol_sx0}
\end{figure}

\begin{figure}
\centering
    \includegraphics[width=0.7\linewidth]{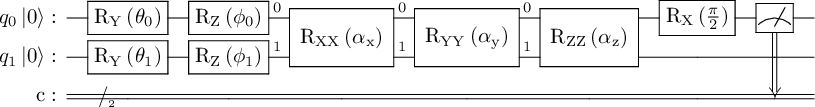}
    \caption{Quantum protocol for calculating the mean value of operator $\sy$}
    \label{fig:protocol_sy0}
\end{figure}

\begin{figure}
\centering
    \includegraphics[width=0.7\linewidth]{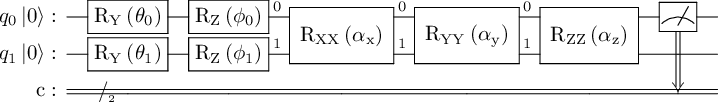}
    \caption{Quantum protocol for calculating the mean value of operator $\sz$}
    \label{fig:protocol_sz0}
\end{figure}

Quantum protocols for the detection of the mean values are presented in figs. \ref{fig:protocol_sx0} - \ref{fig:protocol_sz0}.

\begin{figure}
\centering
    \includegraphics[width=0.5\linewidth]{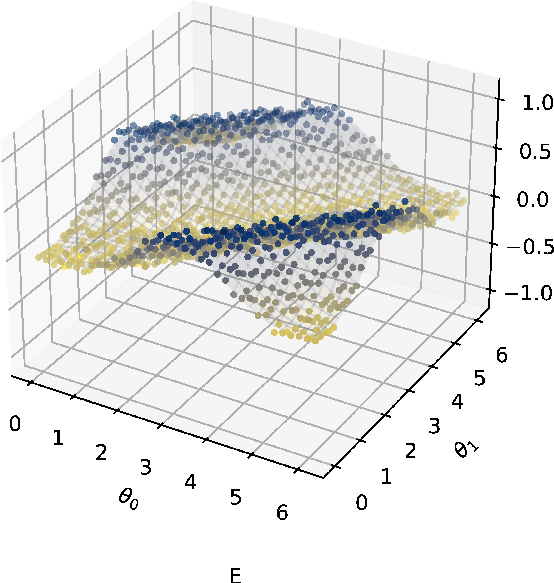}
    \caption{Entanglement distance of qubit $q[0]$ with other qubits in state \eqref{eq:evolution_state} for different values of $\theta_0$ and $\theta_1$.  
    The surface corresponds to the analytical result. Results of quantifying on AerSimulator are marked with dots.}
    \label{fig:dep_theta}
\end{figure}

\begin{figure}
\centering
    \includegraphics[width=0.5\linewidth]{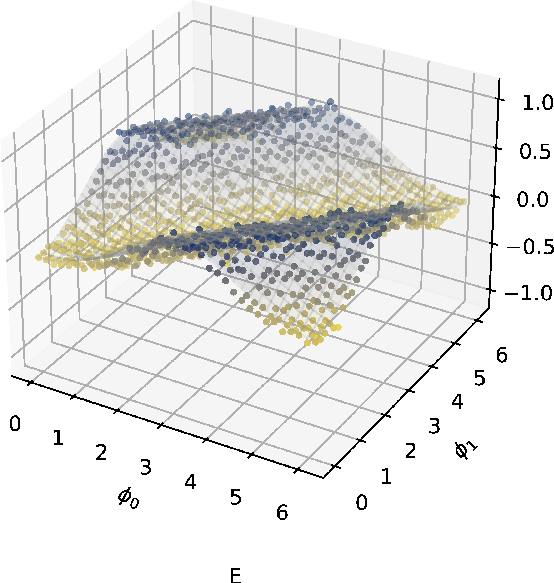} 
    \caption{Entanglement distance of qubit $q[0]$ with other qubits in state \eqref{eq:evolution_state} for different values of $\phi_0$ and $\phi_1$.  The surface corresponds to the analytical result. Results of quantifying on AerSimulator are marked with dots.}
    \label{fig:dep_phi}
\end{figure}

\begin{figure}
\centering
    \includegraphics[width=0.5\linewidth]{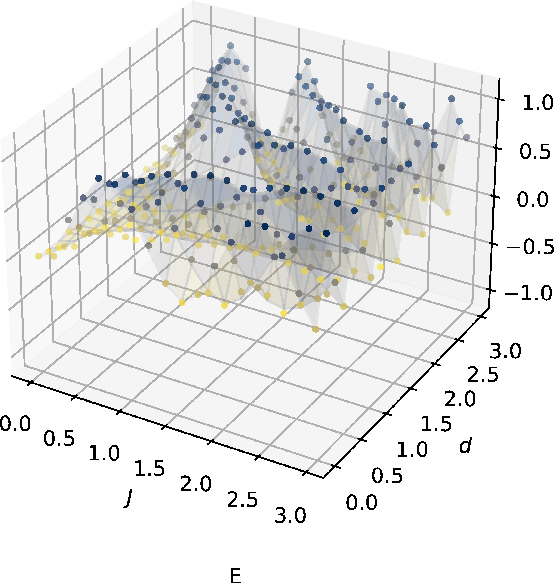} 
    \caption{Entanglement distance of qubit $q[0]$ with other qubits in state \eqref{eq:evolution_state} for different values of $J$ and $d$. The surface corresponds to the analytical result. Results of quantifying on AerSimulator are marked with dots.}
    \label{fig:dep_Jd}
\end{figure}

The entanglement distance for qubit $q[0]$ is obtained for different values of the initial state parameters $\theta_0, \theta_1 \in [0, 2\pi]$ (see fig. \ref{fig:dep_theta}), $\phi_0, \phi_1 \in [0, 2\pi]$(see fig. \ref{fig:dep_phi}), and coupling constants of the model $J, d \in [0, \pi]$ (see fig. \ref{fig:dep_Jd}). We show that the results of quantum computations are consistent with previously derived analytical expressions \eqref{eq:mean_z}, \eqref{eq:mean_x}, \eqref{eq:mean_y}, \eqref{eq:ent_dist}.

The fig. \ref{fig:dep_theta} shows the results of quantum and analytical calculations of the entanglement distance for qubit $q[0]$ with the other qubit. These graphs demonstrate the case when the values of $\phi_0$ and $\phi_1$ and the coupling model parameters are fixed, while the angles $\theta_0$ and $\theta_1$ vary from $0$ to $2\pi$ with steps $\pi/18$.

The figure fig. \ref{fig:dep_phi} shows quantum and analytical calculations of the entanglement distance of qubit $q[0]$ with the other qubit. These graphs demonstrate the case when the values of $\theta_0$ and $\theta_1$ and the coupling model parameters are fixed, while the angles $\phi_0$ and $\phi_1$ vary from $0$ to $2\pi$ with steps $\pi/18$.

The figure fig. \ref{fig:dep_Jd} shows quantum and analytical calculations of the entanglement distance of qubit $q[0]$ with the other qubit. These graphs demonstrate the case where the values of $\theta_0$, $\theta_1$, $\phi_0$ and $\phi_1$ are fixed, while model parameters $J$ and $d$ vary from $0$ to $\pi$ with steps $\pi/18$.

\section{Speed of evolution of quantum states of a two-spin system with the XXZ model}

Let us determine the speed of evolution of the quantum system. According to equation \eqref{eq:speed}, this requires first obtaining $\mean{\DH^2}$.

The quantity $\mean{\DH^2}$ can be evaluated by analyzing the mean value of the evolution operator in short time intervals. In particular, for sufficiently small times, the following approximation can be applied

\begin{multline}
    |S|^2 = |\langle \psi_0(\theta_0, \theta_1, \phi_0, \phi_1)|e^{-iHt/\hbar}|\psi_0(\theta_0, \theta_1, \phi_0, \phi_1) \rangle|^2 \\
    = 1 - \frac{t^2}{\hbar^2} \langle \psi_0(\theta_0, \theta_1, \phi_0, \phi_1)|\DH^2|\psi_0(\theta_0, \theta_1, \phi_0, \phi_1) \rangle.
\end{multline}

\begin{figure}
\centering
    \includegraphics[width=0.8\linewidth]{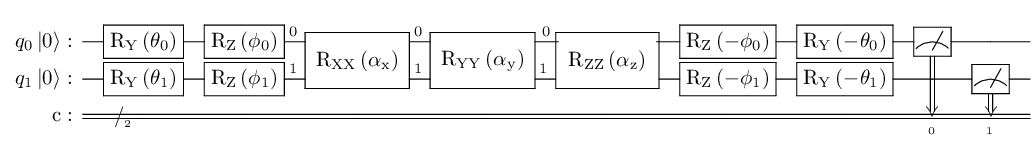}
    \caption{Quantum protocol for calculation $|S|^2$}
    \label{fig:mean_S}
\end{figure}

Quantum protocols for the detection of mean values are presented in fig. \ref{fig:mean_S}.

The quantum protocol represents the expression
\begin{multline}
    |\langle \psi_0(\theta_0, \theta_1, \phi_0, \phi_1)|S|\psi_0(\theta_0, \theta_1, \phi_0, \phi_1) \rangle|^2 \\
    = |\bra{00} RY_0(\theta_0) RY_1(\theta_1) RZ_0(\phi_0) RZ_1(\phi_1) S \\
    \times RZ_0(-\phi_0) RZ_1(-\phi_1) RY_0(\theta_0) RY_1(\theta_1)\ket{00}|^2
\end{multline}

Quantum protocol presented in fig. \ref{fig:mean_S} has been run 1024 times for $\theta_0 = \theta_1 = \pi/2$, $\phi_0 = \phi_1 = \pi/4$, and $d=2$ and for $\alpha_x = \alpha_y = \frac{1}{d} \alpha_z = \alpha$ changing from $-3 \pi/ 32$ to $3\pi / 32$ with step $\pi/128$.

\begin{figure}
    \centering
    \includegraphics[width=0.6\linewidth]{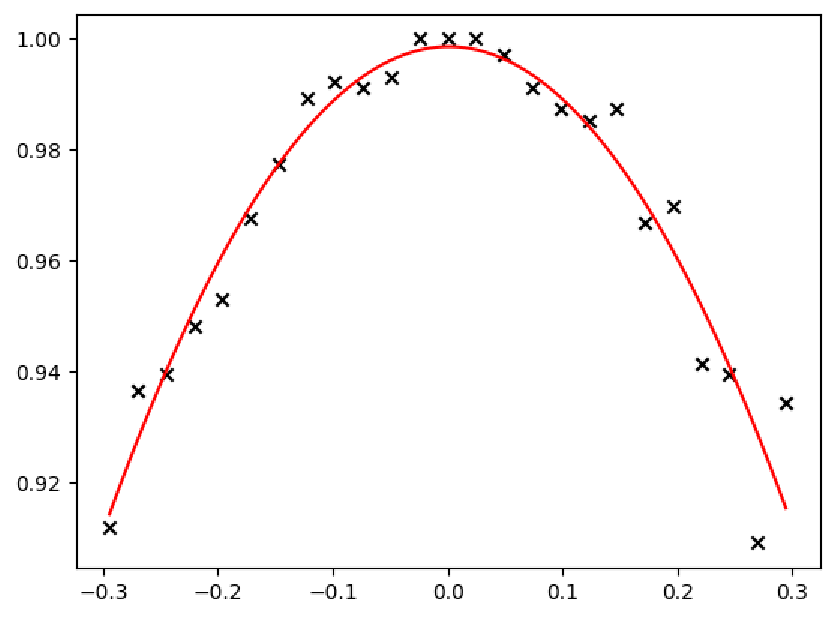}    \caption{Results of quantum calculations for $|S|^2$ as a function of $\alpha = 2tJ/\hbar$ . Marked with cross points ﬁtted by the curve $|S|^2 = -0.96 \alpha^2 + 1$ marked with a line. Although for higher orders of time in the Taylor series, such an approximation may become inaccurate, so higher orders of corrections will be required.}
    \label{fig:result_S2}
\end{figure}

Based on the analysis of the time dependence of $|S|^2$ obtained from quantum calculations, we find $\mean{\DH^2} = 0.96 J^2$. The analytical expression gives $\mean{\DH^2} = J^2$. For the speed of evolution, we obtain $v = 0.98 \gamma |J| / \hbar$, while the theoretical value is $v = \gamma |J| / \hbar$. Thus, the results obtained using the AerSimulator are in close agreement with the theoretical predictions.

We also utilized another approach to quantum calculations of the speed of evolution. 
We consider a two-spin system with XXZ model described by Hamiltonian \eqref{eq:Hamiltonian}. Then,
\begin{equation}
    \mean{H^2} = 2J^2 + J^2 d^2 - 2J^2 \mean{\sz_0 \sz_1}  - 2J^2 d \mean{\sx_0 \sx_1 + \sy_0 \sy_1},
    \label{eq:H2}
\end{equation}
and 
\begin{equation}
    \mean{H}^2 = J^2 \mean{\sx_0 \sx_1 + \sy_0 \sy_1 + d \sz_0 \sz_1}^2.
    \label{eq:meanH2}
\end{equation}
Given that $\mean{\DH^2}=\mean{H^2} - \mean{H}^2$. Therefore, $\mean{\DH^2}$  takes the following form:
\begin{multline}
    \mean{\DH^2} = 2J^2 + J^2 d^2 - 2J^2 \cos{\theta_0}\cos{\theta_1}
    - 2J^2 d \sin{\theta_0}\sin{\theta_1}\cos{(\phi_0 - \phi_1)} \\
    - J^2 \sin^2\theta_0 \sin^2\theta_1 \cos^2(\phi_0 - \phi_1)
    - \frac{1}{2} J^2 d \sin 2\theta_0 \sin 2\theta_1 \cos(\phi_0 - \phi_1) \\
    - J^2 d^2 \cos^2 \theta_0 \cos^2 \theta_1
\end{multline}

So, based on this result for the speed of quantum evolution, we have

\begin{multline}
    v = \frac{\gamma}{\hbar} \bigl(2J^2 + J^2 d^2 - 2J^2 \cos{\theta_0}\cos{\theta_1}
    - 2J^2 d \sin{\theta_0}\sin{\theta_1}\cos{(\phi_0 - \phi_1)}\\
    - J^2 \sin^2\theta_0 \sin^2\theta_1 \cos^2(\phi_0 - \phi_1)
    - \frac{1}{2} J^2 d \sin 2\theta_0 \sin 2\theta_1 \cos(\phi_0 - \phi_1)\\
    - J^2 d^2 \cos^2 \theta_0 \cos^2 \theta_1 \bigr)^{1/2}.
\end{multline}

According to results \eqref{eq:H2} - \eqref{eq:meanH2}, the speed of evolution can be obtained based on the computations of $\mean{\sigma_0^k \sigma_1^k}$, $k=x,y,z$.

\begin{equation}
    \mean{\sx_0\sx_1} = \mean{\psi(\theta_0, \theta_1, \phi_0, \phi_1) |\sx_0\sx_1|\psi(\theta_0, \theta_1, \phi_0, \phi_1)}
\end{equation}

\begin{equation}
    \mean{\sy_0\sy_1} = \mean{\psi(\theta_0, \theta_1, \phi_0, \phi_1) |\sy_0\sy_1|\psi(\theta_0, \theta_1, \phi_0, \phi_1)}
\end{equation}

\begin{multline}
    \mean{\sz_0\sz_1} = \mean{\psi(\theta_0, \theta_1, \phi_0, \phi_1) |\sz_0\sz_1|\psi(\theta_0, \theta_1, \phi_0, \phi_1)}  \\ 
    =
    |\langle 00 | \psi_0(\theta_0, \theta_1, \phi_0, \phi_1) \rangle|^2
    - |\langle 01 | \psi_0(\theta_0, \theta_1, \phi_0, \phi_1) \rangle|^2
     \\
    - |\langle 10 | \psi_0(\theta_0, \theta_1, \phi_0, \phi_1) \rangle|^2
    + |\langle 11 | \psi_0(\theta_0, \theta_1, \phi_0, \phi_1) \rangle|^2
\end{multline}

\begin{figure}
\centering
    \includegraphics[width=0.55\linewidth]{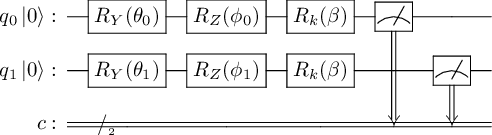}
    \caption{Quantum protocol for calculating the mean value of operators $\sx_0\sx_1$, $\sy_0\sy_1$, and $\sz_0\sz_1$}
    \label{fig:protocol_s0s1}
\end{figure}

Quantum protocols for such studies are presented in fig. \ref{fig:protocol_s0s1}.

\begin{figure}
\centering
    \begin{minipage}[t]{0.32\linewidth}
        \centering
        \includegraphics[width=\linewidth]{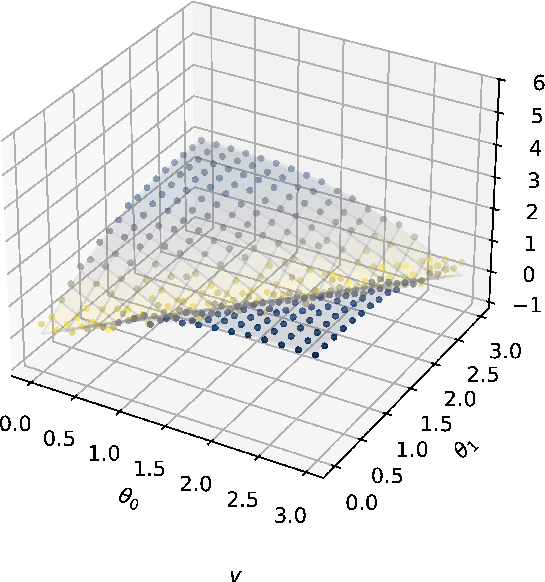}
        \small (a)
    \end{minipage}\hfill
    \begin{minipage}[t]{0.32\linewidth}
        \centering
        \includegraphics[width=\linewidth]{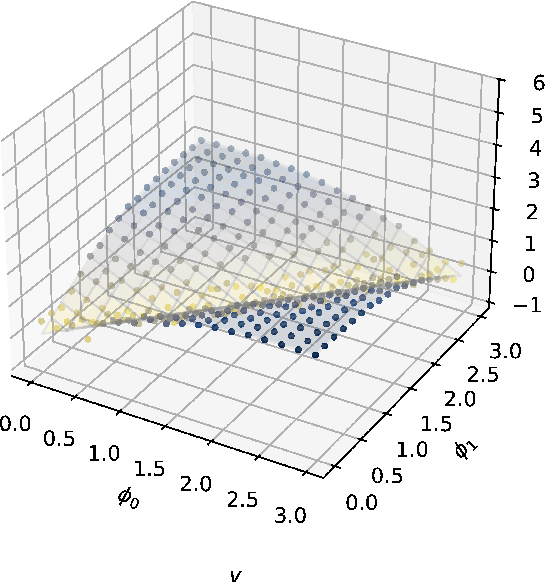}
        \small (b)
    \end{minipage}\hfill
    \begin{minipage}[t]{0.32\linewidth}
        \centering
        \includegraphics[width=\linewidth]{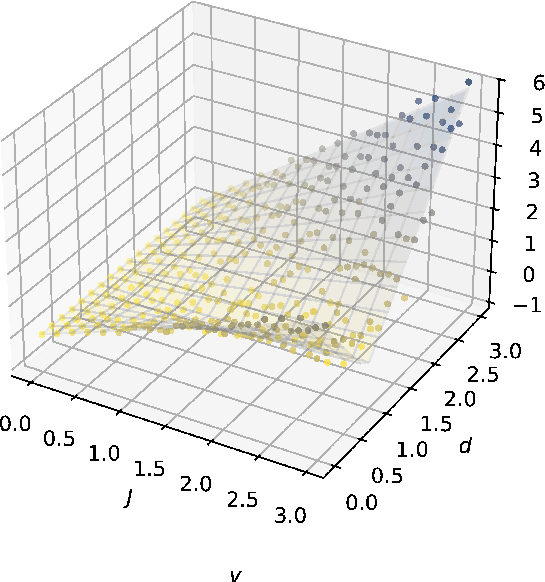}
        \small (c)
    \end{minipage}

    \caption{Results of quantum calculations of the speed of evolution on the basis of quantifying $\mean{\sigma_0^k \sigma_1^k}$, $k = x, y, z$ represented by dots. The solid surface represents the results obtained analytically.
    (a) The values of fixed parameters $\phi_0 = \phi_1 = \pi/4$, $J=1$, $d=1$. (b) The values of fixed parameters $\theta_0 = \theta_1 = \pi/2$, $J=1$, $d=1$. (c) The values of fixed parameters $\theta_0 = \theta_1 = \pi/2$, $\phi_0 = \phi_1 = \pi/4$.}
    \label{fig:speed_res}
\end{figure}

The results of the studies are shown in fig. \ref{fig:speed_res}. 
The graph (a) demonstrates the case where the values of $\phi_0$, $\phi_1$, and the coupling model parameters are fixed, while the values of $\theta_0$, $\theta_1$ vary from $0$ to $\pi$ with steps $\pi/18$.
The graph (b)demonstrates the case where the values of $\theta_0$, $\theta_1$, and the coupling model parameters are fixed, while the values of $\phi_0$ and $\phi_1$ vary from $0$ to $\pi$ with steps $\pi/18$.
The graph (c) shows the case where the values of $\theta_0$, $\theta_1$, $\phi_0$, and $\phi_1$ are fixed, while the model parameters $J$ and $d$ vary from $0$ to $\pi$ with steps $\pi/18$.

\section{ Conclusions} 
The two-spin system with the XXZ model has been studied. We have examined the entanglement distance of the quantum states of the system both analytically and using quantum computing.
We have considered two-qubit states given by \eqref{eq:init_state}, \eqref{eq:evolution_state}.
Relations between entanglement and parameters of an arbitrary quantum state have been presented. 
An expression for entanglement has been obtained for the case in which the initial state is an arbitrary separable state of two spins (see \eqref{eq:init_state}). 
The dependence of the entanglement on the values of the coupling constants of the model and the parameters of the initial states has been found analytically
(\eqref{eq:mean_z}, \eqref{eq:mean_x}, \eqref{eq:mean_y}, \eqref{eq:ent_dist}).

In addition, the entanglement of a two-qubit state has been studied on the basis of quantum computations.
Quantum protocols for quantifying entanglement have been constructed (figs. \ref{fig:protocol_sx0} - \ref{fig:protocol_sz0}) and run on AerSimulator. We have determined the entanglement of the two-spin state based on the study of the mean value of the evolution operator.

Results of quantum calculations for mean values and entanglement are in good agreement with the theoretical ones (see fig. \ref{fig:dep_theta} - \ref{fig:dep_Jd}).

The speed of evolution of the system and its dependence on the parameters of the initial state and the model parameters were studied.
An analytical expression for the speed of evolution was obtained for the case of an arbitrary separable two-qubit initial state \eqref{eq:init_state}. We have determined the speed of evolution based on the study of the mean value $\mean{\DH^2}$.
To determine $\mean{\DH^2}$, for various values of initial state parameters $\theta_0$, $\theta_1$, $\phi_0$, $\phi_1$ as well as coupling model parameter $J$ and $d$, the $\mean{\sigma_0^k \sigma_1^k}$, $k=x, y, z$ was calculated.
The dependence of the speed of evolution on the parameters of the initial state and model parameters has been investigated both analytically and through quantum computing (see fig. \ref{fig:speed_res}). The theoretical predictions are in good agreement with the results of quantum computations.

\end{document}